\documentclass[prb,aps,twocolumn,floatfix,amsmath,amssymb,tightenlines, superscriptaddress]{revtex4}
\usepackage[pdftex]{graphicx}
 \usepackage{amsmath}
\usepackage{flushend}
\usepackage[utf8x]{inputenc}
\usepackage[T1]{fontenc}
\usepackage{amssymb}
\usepackage{amsfonts}
\usepackage{bm}
 \usepackage{amsmath} 
 \usepackage[breaklinks=true,colorlinks=true,linkcolor=blue,urlcolor=blue,citecolor=blue]{hyperref}
\usepackage{subfigure}
\usepackage{lipsum}
\usepackage{amsfonts} 
\usepackage{amssymb, mathrsfs}
\usepackage{braket}
\usepackage{graphicx} 
\usepackage[usenames]{color} 

\usepackage{bbm}
\def\beq{\begin{equation}}
\def\eeq{\end{equation}}
\def\bsp{\begin{split}}
\def\esp{\end{split}}
\def\bea{\begin{eqnarray}}
\def\eea{\end{eqnarray}}
\def\ba{\begin{array}}
\def\ea{\end{array}}

\def\lb{\left(}
\def\rb{\right)}

\def\l.{\left.}
\def\r.{\right.}

\begin{document}

\title{Weyl semimetal in ultra-thin film of topological insulator multilayer}
\author{S. A. Owerre}
\email{sowerre@perimeterinstitute.ca}
\affiliation{African Institute for Mathematical Sciences, 6 Melrose Road, Muizenberg, Cape Town 7945, South Africa.}
\affiliation{Perimeter Institute for Theoretical Physics, 31 Caroline St. N., Waterloo, Ontario N2L 2Y5, Canada.}
\begin{abstract}
In ultra-thin film of topological insulator,  the hybridization  between the top and bottom surfaces   opens an energy gap and forms two degenerate quantum anomalous Hall states, which give rise to  a quantum spin Hall state.   In this paper, we demonstrate  that a  Weyl semimetal can be realized in an ultra-thin film of topological insulator  heterostructure in a similar way to that of the surface state of a strong three-dimensional (3D) topological insulator studied by Burkov and Balents . We find that the system realizes  both 3D quantum anomalous Hall phase and  3D quantum spin Hall phase, and the Weyl nodes occur at zero energy when both time-reversal symmetry and inversion symmetry are explicitly broken by the magnetic field and the structure inversion asymmetry of the thin film.
\end{abstract}

\maketitle
%

\section {Introduction} 

Weyl semimetal is a name given to specific materials that host Weyl fermions in three dimensions in which time-reversal symmetry ($\mathcal{T}$)   or inversion symmetry ($\mathcal{I}$)  is explicitly broken \cite{aab,aab0,gab}. The low-energy Hamiltonian near the isolated band touching points (Weyl points) is reminiscent of the Weyl equation or massless Dirac equation  from high energy physics \cite{mur,wan}, given by $H=\pm v_F\boldsymbol{\sigma}\cdot\bold{k}$, where $\boldsymbol{\sigma}$ is the triplet Pauli matrices and $\bold{k}$ is a 3-component Brillouin zone momentum and $\pm$ denote the chirality. In contrast to two-dimensional (2D) electron systems \cite{ zhang, hk}, the Weyl point is robust to any external perturbations as the three components of the momentum are required to vanish at the point of degeneracy. The breaking of $\mathcal{T}$ or $\mathcal{I}$ separates the Weyl nodes in momentum space and they become topologically stable\cite{vol}.  The topological properties of Weyl points  are manifested as monopoles of Berry flux in the the Brillouin zone (BZ) with  point-like Fermi arcs surface states \cite{wan}. 

However, when the Fermi arcs are not point-like but  tilted,  there is another possibility of   Weyl semimetals dubbed type-II Weyl semimetals\cite{solu}.  In recent years, there have been many proposals of systems that host  Weyl semimetal phases. The pyrochlore iridates\cite{wan, kre} are well-known as one of the materials in which a Weyl semimetal phase can be observed. A very simple model using the surface states of a strong 3D TI heterostructure has been proposed\cite{aab,aab1,aab2,aab0}.  Weyl semimetal phase  has been proposed in the magnetically doped topological band insulators\cite{liu}. There is also a toy tight binding model proposal that captures the existence  of Weyl semimetal phase \cite{hui5, hui6,hui7}. Recently, Weyl semimetal has been discovered experimentally in photonic crystals\cite{llu}.  The experimental realization of  Weyl semimetal in TaAs has also been reported  using angle-resolved photoemission spectroscopy \cite{Xu,lv, lv1}.


In this letter, we demonstrate that  Weyl semimetal can be realized  in  an alternating ordinary insulator and thin film of topological insulator (TI)  multilayer.  This heterostructure has some similarities to the model studied by Burkov and Balents \cite{aab}. However, the ultrathin film of TI possesses a different Hamiltonian \cite{hui,hui1,hai1, hui2}. Besides, they  can be easily grown in the laboratory and have been realized in most experiments in ultrathin Bi$_2$Se$_3$ and Bi$_2$Te$_3$ films \cite{zza1, hui3, hui4}. In experiments,  it is indeed possible to grow quintuple layers (QL) of an ultrathin film of TI, in which two Bi and three Se or Te layers are stacked together.  

 In particular,  we show that in the absence of a magnetic field and structure inversion asymmetry the thin film of TI multilayer captures  a phase with two Dirac nodes, which annihilate each other at the phase transition point and opens a gap to a 3D QSH phase.  In the presence of an external magnetic field, however, the system breaks $\mathcal{T}$ but preserves $\mathcal{I}$. We find that the magnetic field introduces a 3D QAH phase. We also find a phase with two Weyl points separated in momentum space (Weyl semimetal) and an ordinary insulator. Adding a structure inversion asymmetry term introduces a potential difference between the top and bottom surfaces of each layer and breaks $\mathcal{I}$ but preserves $\mathcal{T}$, while the magnetic field breaks $\mathcal{T}$ but preserves $\mathcal{I}$, and the whole system then breaks both symmetries. In this case, the system still produces a Weyl semimetal phase and the nodes still occur at zero energy.  We also analyze the effects of an orbital magnetic field and obtain the Landau level spectra of the system. The zero Landau levels capture the appearance of Weyl semimetal phase in the vicinity of the bulk gap.  Finally, we show that in the absence of the magnetic field and the structure inversion asymmetry,  the continuum limit of the ultrathin film TI multilayer is the same as the continuum limit of the toy tight binding Hamiltonian of Weyl semimetal\cite{hui5, hui6,hui7, hui8}.

\section { Model}
\begin{figure}[ht]
\centering
\includegraphics[width=1.8in]{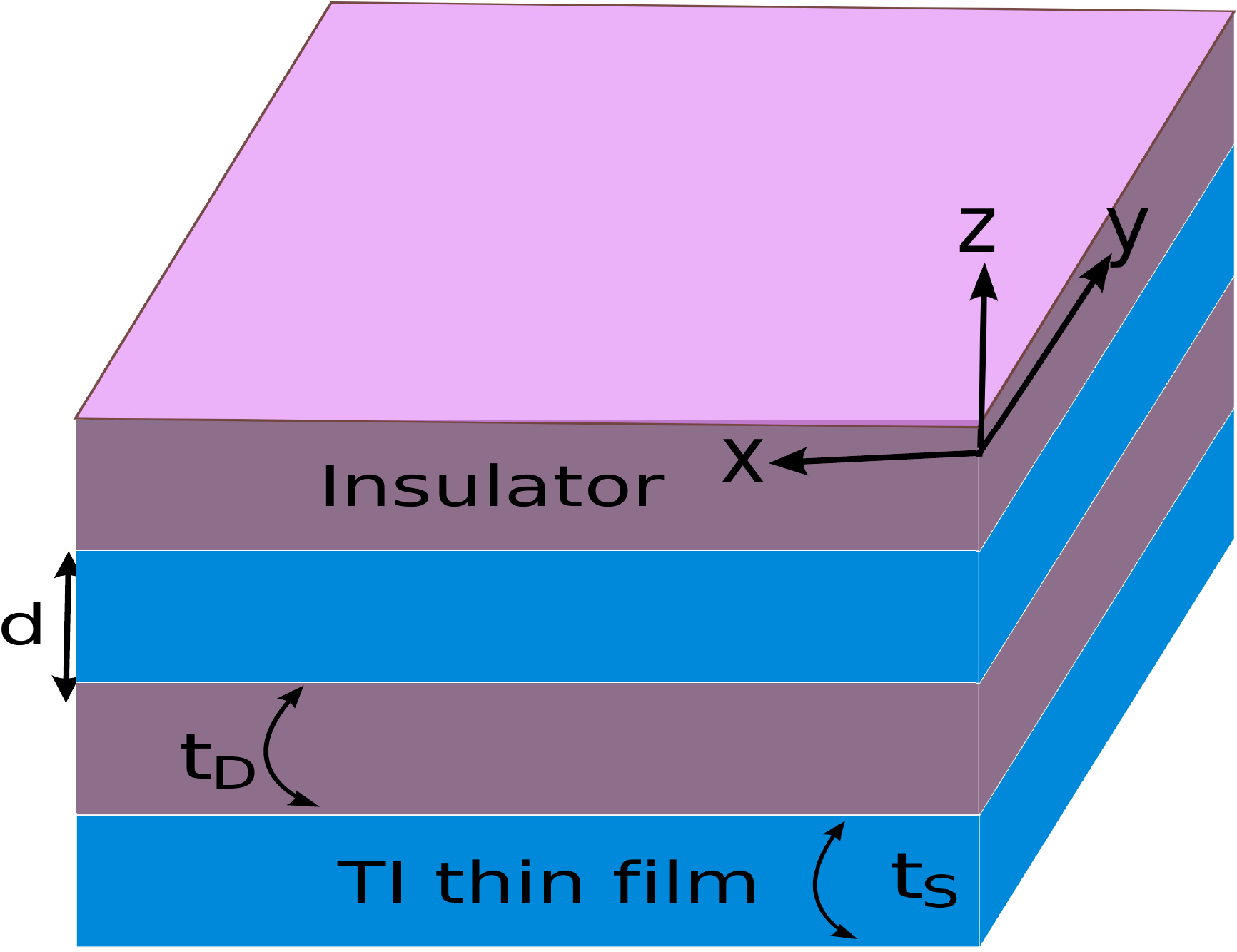}
\caption{Color online. Schematic sketch of four array of alternating ordinary insulator and thin film topological insulators (TI), where  $t_S$ is the hybridization potential that couples the top and bottom surfaces in the same thin film layer, $t_D$  is the hybridization potential on neighbouring thin film layers, $d$ is the spacing between film layers. }
\label{hetero}
\end{figure}

 Following Burkov and Balents \cite{aab}, we study a simple Hamiltonian of an ultrathin film of topological insulator heterostructure (Fig.~\ref{hetero}). The Hamiltonian is governed by 

\begin{align}
H&= \sum_{\bold{k}_\perp,ij}a^\dagger_{\bold{k}_\perp i}\mathcal{H}_{ij}a_{\bold{k}_\perp j},
 \end{align}
 where
 \begin{align}
\mathcal{H}_{ij}&= v_F(\hat{z}\times \boldsymbol{\sigma})\cdot\bold{k}_\perp\delta_{ij} + (\frac{t_S}{2}-t_\perp k_\perp^2)\tau_z\sigma_z\delta_{ij} \label{genbhz} \nonumber\\& +  b\tau_{x}\delta_{ij}+\gamma\sigma_z\delta_{ij} + \frac{t_D}{2}\frac{(\delta_{j,i+1}+\delta_{j,i-1})}{2}\tau_z\sigma_z .
 \end{align}
Here,  $\boldsymbol{\sigma}$ are the Pauli matrices  on the real spin space and $\boldsymbol{\tau}$ are the {\it which surface} pseudo spins. $\bold{k}_\perp=(k_x,k_y)$ is a 2D momentum vector in the BZ. The indices $i,j$ label distinct thin film layers and $v_F$ is the Fermi velocity. 
 
 The first two terms in Eq.~\eqref{genbhz} describe the low-energy Hamiltonian of a single 2D ultrathin film of TI layer \cite{hui,hui1,hai1, hui2}. The parameters $t_S$  and $t_\perp$ are the hybridization potentials that couple the top and bottom surfaces of the same thin film layer for small $k_\perp$ and large $k_\perp$ respectively. The parameter  $b$ denotes the structure inversion asymmetry term which will be inevitably present when growing the system in Fig.~\eqref{hetero}. In fact, the first three terms in Eq.~\eqref{genbhz} have recently been studied\cite{luz}. The electrostatic potential    introduces a potential difference of  $2b$ between the top and bottom surfaces.   The Zeeman splitting is $\gamma=g\mu_B B$ and can be induced by depositing a ferromagnetic material on the thin film (magnetic doping) or directly applying a magnetic field. Here $g$ is the Land\'e $g$-factor of the thin film, $\mu_B$ is the Bohr magneton, and $B$ is the magnetic field. The new parameter we introduce in Eq.~\eqref{genbhz}   is  $t_D$. This is the  hybridization potential that couples the top and bottom surfaces of neighbouring thin film layers along the growth $z$-direction.
 
  The parameters  $b$, $t_\perp$, $t_S$, and $t_D$ depend on the thickness of the thin film. The first three parameters have been determined both  numerically \cite{hui,hui1,hui2} and experimentally\cite{hui4,zza1}. The new parameter $t_D$ can also be determined by growing the multilayer in Fig.~\eqref{hetero}. Without loss of generality we assume all the parameters to be positive $b, t_\perp, t_S,t_D>0$, which is indeed the case in most experiments in 2D thin films $(t_D=0)$\cite{zza1, hui4}.  

%

 \section{Zero field and zero potential}
\label{zerofp}
 Let us first consider a very simple  case $\gamma=b=0$ in Eq.~\eqref{genbhz}. To study the topological properties of this system it is expedient to Fourier transform into momentum space along the growth  $z$-direction. We obtain $H=\sum_{\bold{k}}a^\dagger_{\bold{k}}\mathcal{H}(\bold{k})a_{\bold{k}}$, with

\begin{align}
\mathcal{H}(\bold{k})&= v_F(\hat{z}\times \boldsymbol{\sigma})\cdot\bold{k}_\perp +  \hat{\Delta}(k_z, k_\perp)\sigma_z,
\label{fullti}
\end{align}
where 
\bea
\hat{\Delta}(k_z, k_\perp)=\lb \frac{t_S}{2}-t_\perp k_\perp^2+ \frac{t_D}{2}\cos(k_zd)\rb \tau_z.
\label{det}
\eea
One distinguishing feature of $\hat{\Delta}$ is that it contains  quadratic terms in the continuum limit, which is in fact a common feature of  toy tight binding models of Weyl semimetal \cite{hui8, hui5}. 
For the model in Eq.~\eqref{fullti},  the system preserves $\mathcal{T}$ and $\mathcal{I}$.  It may be written as a four-component Dirac fermions
 \begin{align}
 \mathcal{H}(\bold{k})=
 \begin{pmatrix}
 \mathcal{H}_\uparrow(\bold{k})&0\\
 0 &\mathcal{H}_\uparrow^*(-\bold{k})
 \end{pmatrix},
 \label{decou}
 \end{align}
 where the arrow denote the top surface and the bottom surface is related by $\mathcal{H}_\uparrow^*(-\bold{k})=\sigma_y\mathcal{H}_\downarrow(\bold{k})\sigma_y$ under  a unitary transformation  using $U=\text{diag}(\sigma_0, i\sigma_y)$, where $\sigma_0$ is a ${2\times 2}$ identity matrix. Under  time-reversal symmetry and inversion symmetry, we have 
 \begin{align}
 &\mathcal{T}:~\mathcal{H} (\bold{k})\to \mathcal{T}\mathcal{H}^*(-\bold{k}) \mathcal{T}^{-1},\\& \mathcal I:~\mathcal{H}(\bold{k})\to \mathcal{I}\mathcal{H}(-\bold{k}) \mathcal{I}^{-1}.
 \end{align}
    The $\mathcal{T}$ operator is  $\mathcal{T}=\tau_0\otimes \Theta$, where $\Theta=i\sigma_y \mathcal{K}$ and   $\mathcal{K}$ is the complex conjugation. The inversion operator for this system is $\mathcal{I}=\tau_z\otimes \sigma_z$.  Note that $\mathcal{T}^2=-1$ and $\mathcal{I}^2=1$.  The system thus describes a 3D Dirac semimetal. The eigenvalues of $\hat{\Delta}(k_z, k_\perp)$ are $\pm\Delta(k_z, k_\perp)$, where
\bea \Delta(k_z, k_\perp)= \frac{t_S}{2}- t_\perp k_\perp^2+ \frac{t_D}{2}\cos(k_zd).\eea The corresponding eigenspinors are
\begin{align}
u^\uparrow=\begin{pmatrix}1\\ 0\end{pmatrix};\quad u^\downarrow=\begin{pmatrix}0\\ 1\end{pmatrix}.
\label{eig0}
\end{align}
 The Hamiltonian then becomes
\begin{align}
\mathcal{H}_s(\bold{k})&= v_F(\hat{z}\times \boldsymbol{\sigma})\cdot\bold{k}_\perp +s\Delta(k_z, k_\perp)\sigma_z,
\label{par}
\end{align}
where $s=\pm~(\uparrow,\downarrow)$. 

Similar to Burkov and Balents \cite{aab} model at zero magnetic field,  $\Delta(k_z, k_\perp)$ vanishes at $t_S/t_D=1$ for $k_z=\pi/d$ and at $t_S/t_D=-1$ for $k_z=0$, only if $k_x=k_y=0$.  The Hamiltonian  is easily diagonalized. The energy eigenvalues are two-fold degenerate as expected from Kramers theorem due to time-reversal symmetry.  They are given by \begin{align}
 \epsilon_{\eta} (\bold k)=\eta\sqrt{v_F^2k_\perp^2+\Delta^2}=\eta\varepsilon_{\bold k}, 
\end{align}
where $\Delta=\Delta(k_z,k_\perp)$ and $\eta=\pm$ labels the conduction and the valence bands respectively, and the eigenvectors are
\bea \chi_\eta(\Delta) = \frac{1}{\sqrt{2}}\lb\sqrt{1-\eta\frac{\Delta}{\varepsilon_{\bold k}}},-i\eta e^{-i\theta_{\bold k_\perp}}\sqrt{1+\eta\frac{\Delta}{\varepsilon_{\bold k}}}\rb^T\label{eig1},\eea where $\theta_{\bold k_\perp} =\text{tan}^{-1}\lb k_y/k_x \rb$.
Hence, the eigenspinors of the complete system are the tensor product of Eqs.~\eqref{eig0} and \eqref{eig1} given by
 \begin{align}
 \psi_\eta^{\uparrow}&=  {\chi_\eta(\Delta) \choose \bold{0}} \quad\text{and} \quad \psi_\eta^{\downarrow}=  {\bold{0}\choose \chi_\eta(-\Delta)}.
 \label{spino}
  \end{align}
  
 For $t_D\neq 0$, the system can be regarded as  two copies of 3D two-band QAH insulators, each breaks $\mathcal{T}$ and the whole system is $\mathcal{T}$-invariant. There are two Dirac nodes  located  along the line $k_x=k_y=0$, $k_z= \pi/d \pm k_z^0$, where
 \bea k_z^0=\frac{1}{d}\arccos\lb\frac{t_S}{t_D}\rb.
 \eea
 \begin{figure}[ht]
\centering
\includegraphics[width=2.8in]{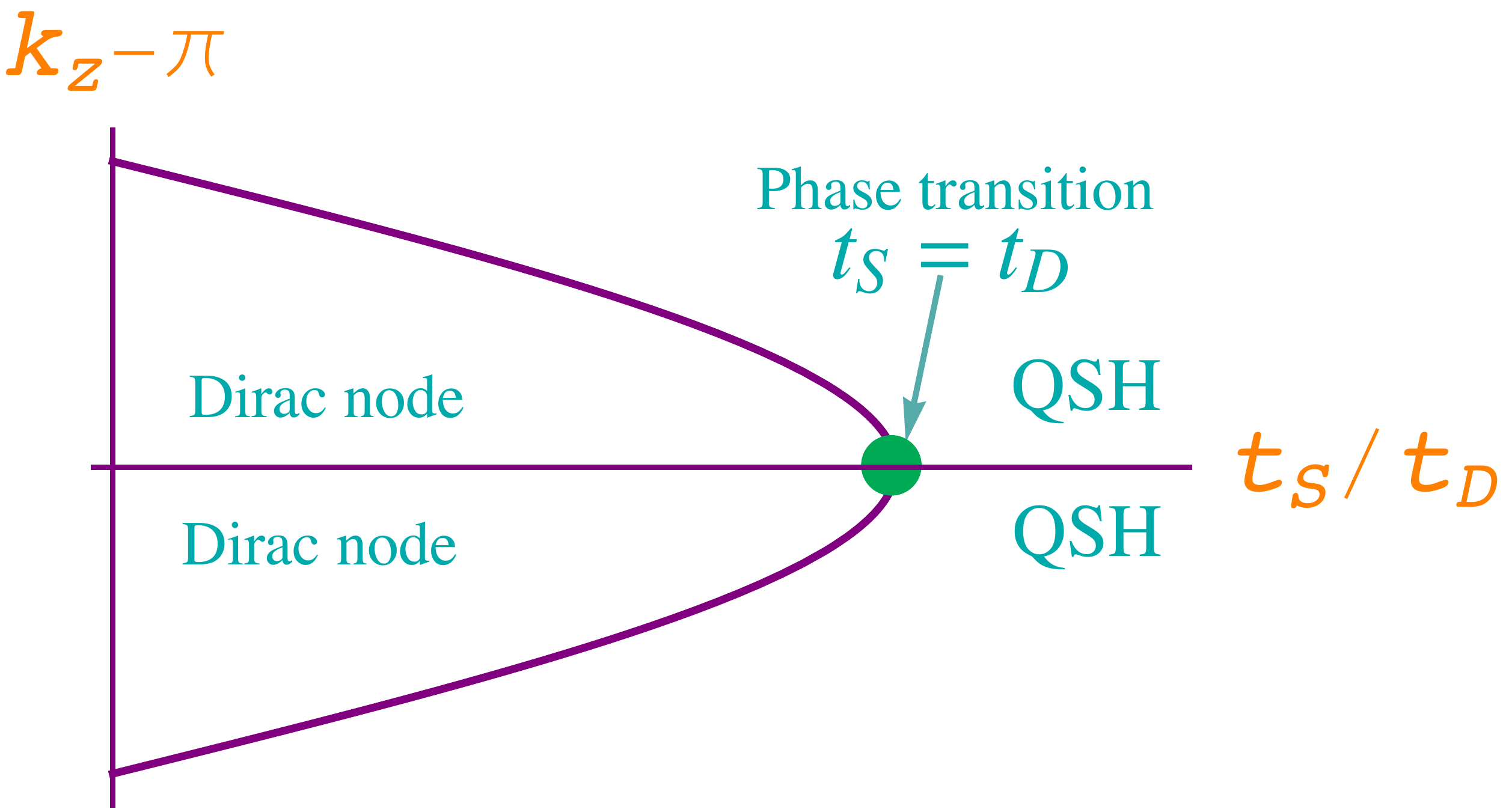}
\caption{Color online. The topological phase transition in $(t_S/t_D,k_z)$ space at $\gamma=b=0$. The two Dirac nodes are not protected by symmetry and they are required to annihilate at the phase transition point $t_S=t_D$ giving rise to fully gapped phase (QSH phase)   for $t_S>t_D$.}
\label{phase0}
\end{figure}
We see that the $\mathcal{T}$-invariant ultrathin film TI multilayer captures  a phase with two Dirac nodes along the $k_z$-axis for $t_S<t_D$.  However, since having only $\mathcal T$ and $\mathcal I$ symmetries cannot produce stable Dirac points with non-trivial topological invariant number \cite{mur, yang},  additional symmetry (e.g. crystal rotational symmetry) is required to stabilize the Dirac points as in  the case of 3D Dirac semimetal   proposed  in Cd$_3$As$_2$ and Na$_3$Bi \cite{wang, wang1,wang2}. Thus, in the present model the Dirac points must annihilate each other (see Fig.~\ref{phase0}). This is what happens at the topological phase transition point  $t_S/t_D=1$. A gap reopens for $t_S>t_D$, which corresponds to a 3D QSH phase.  The  topological invariant number, $\nu$, in the vicinity of the Dirac points must vanish by time-reversal symmetry\cite{mur,yang}. In the 3D QSH phase ($t_S>t_D$),  we compute the $Z_2$ topological number by diagonalizing the parity operator in the occupied bands    at the phase transition point $k_x=k_y=0,~k_z=\pi/d$, using the method of Fu and Kane \cite{fuk1}. We obtain
\bea
(-1)^\nu= -\text{sgn}(t_S-t_D).
\eea
 \section{Nonzero field and zero potential}
 When a nonzero magnetic field is applied to the system, the 3D Dirac semimetal can be driven into various topologically distinct phases since time-reversal symmetry is broken\cite{wang1, aab}. Keeping  $b=0$ preserves inversion symmetry.  In this case, the resulting Hamiltonian is given by 
\begin{align}
{\mathcal{H}}_s(\bold{k})&= v_F(\hat{z}\times \boldsymbol{\sigma})\cdot\bold{k}_\perp + m_s(k_z,k_\perp)\sigma_z,
\label{transti}
\end{align}
where $m_s(k_z,k_\perp)=\gamma + s \Delta (k_z,k_\perp)$. 
The above Hamiltonian  (Eq.~\ref{transti}) is very similar to that of Burkov and Balents \cite{aab} except for the ${\Delta}(k_z,k_\perp)$ function.  The effect of breaking time-reversal symmetry is that the degeneracy of each Dirac node split into two Weyl nodes (the kissing of two non-degenerate bands) separated in momentum space.  This is clearly seen by diagonalizing the Hamiltonian with $\gamma\neq 0$. The eigenvalues  are given by
\begin{align}
\epsilon_{\eta s} (\bold k) =\eta\varepsilon_{s} (\bold k)=\eta\sqrt{v_F^2k_\perp^2+m_s^2(k_z,k_\perp)}.
\end{align} 
 The eigenspinors are given by Eq.~\eqref{spino} with the replacement 
 $\varepsilon(\bold{k})\to \varepsilon_\pm(\bold{k})$; $\pm\Delta(k_z,k_\perp)\to m_\pm(k_z,k_\perp)$. Now the Dirac nodes  are given by the solutions of $m_-(k_z,k_\perp)=0$. Note that the positive mass never changes sign.  The  Dirac nodes  are located at  $k_x=k_y=0$, $k_z= \pi/d \pm k_z^0$, where
\begin{align}
k_z^0=\frac{1}{d}\arccos\lb 1-\frac{2}{t_D}[\gamma-\frac{t_S-t_D}{2}]\rb.
\label{nod1}
\end{align}
We denote the phase boundaries by  $\gamma_\pm=(t_S\pm t_D)/2$, where $\gamma_+ >|\gamma_-|$.   The phase diagram comprises an ordinary insulator phase for $\gamma<|\gamma_-|$ and  a 3D Weyl semimetal phase in the regime $|\gamma_-|<\gamma<\gamma_+$, with two Dirac nodes. For  $\gamma>\gamma_+$ the system transits into a 3D QAH phase. 
The Chern number depends only on the $\tau_z=-1$ sector. It can be computed  using the eigenstates of the occupied bands in the $\tau_z=-1$ sector, we obtain\cite{fuk}

\begin{equation}
 \mathscr{C}^{-}(k_z)=-\frac{1}{2}[\text{sgn}\lb \gamma-\Delta(k_z,0)\rb+\text{sgn}(t_\perp)].
 \label{chern}
\end{equation}
The total conductivity is quantized at $k_z^0=\pi/d$ or $k_z=0$. It is given by
\begin{align}
\sigma_{xy}= -\frac{e}{2hd}[\text{sign}\lb \gamma-\gamma_+\rb+\text{sign}(t_\perp)].
\end{align}

 \section{Nonzero field and nonzero potential}
 Now we study the effects of breaking both time-reversal symmetry and inversion symmetry. This can be achieved by turning on the structure inversion asymmetry, $b\neq0$.  The Hamiltonian  is given by 
 \begin{align}
 \mathcal{H}(\bold{k})&= v_F(\hat{z}\times \boldsymbol{\sigma})\cdot\bold{k}_\perp +  [\gamma+\hat{\Delta}(k_z, k_\perp)]\sigma_z +b\tau_x.
\label{fullti1}
\end{align}
 \begin{figure}[ht]
\centering
\includegraphics[width=3in]{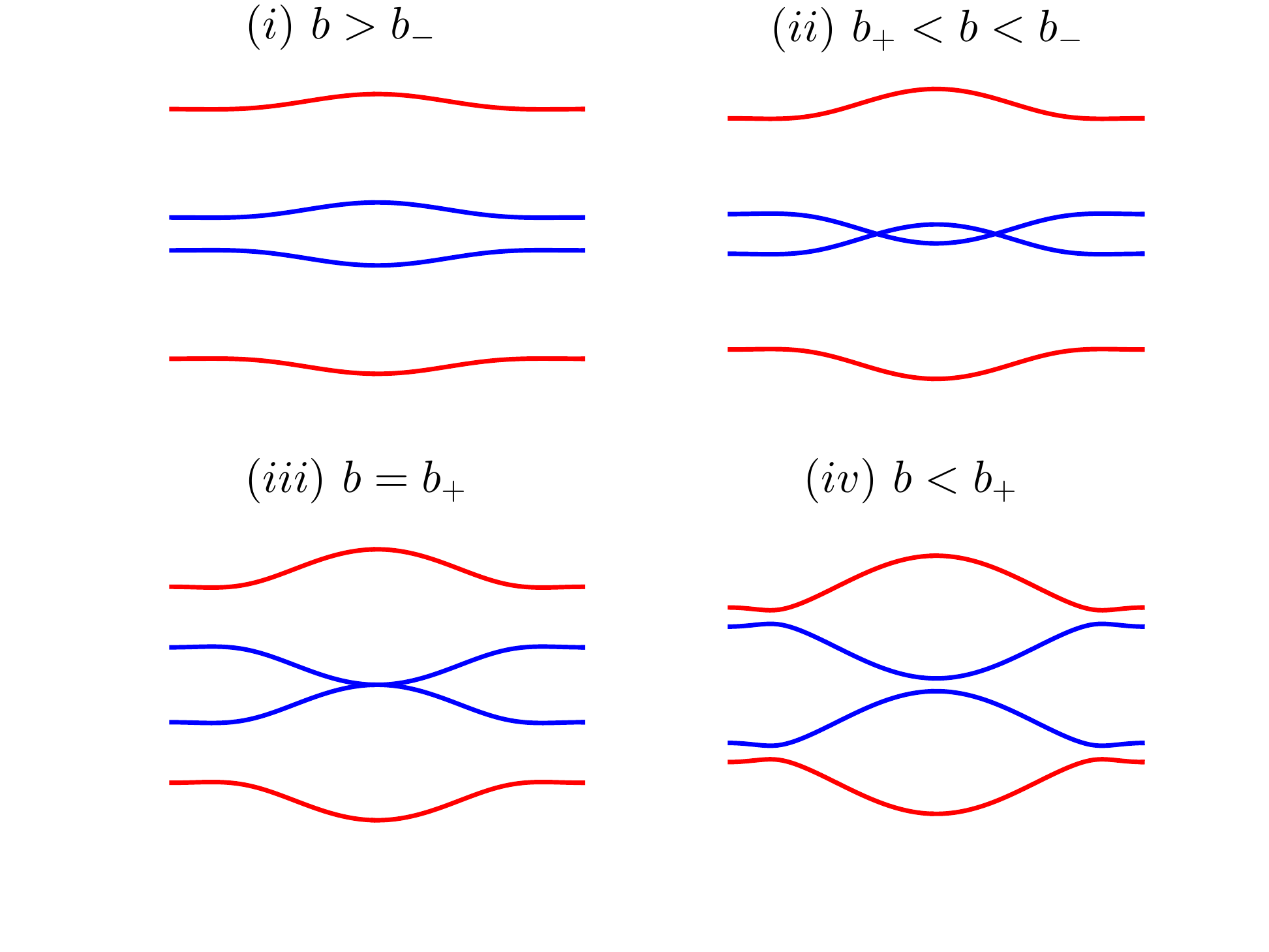}
\caption{Color online. The energy dispersion of Eq.~\eqref{fullti} along the growth direction, that is $k_z$ direction in momentum space with $k_x=k_y=0$. We take  $t_S<t_D$, $b<\gamma$, and $\gamma>\gamma_\pm$.}
\label{phase}
\end{figure}
\begin{figure}[ht]
\centering
\includegraphics[width=2.8in]{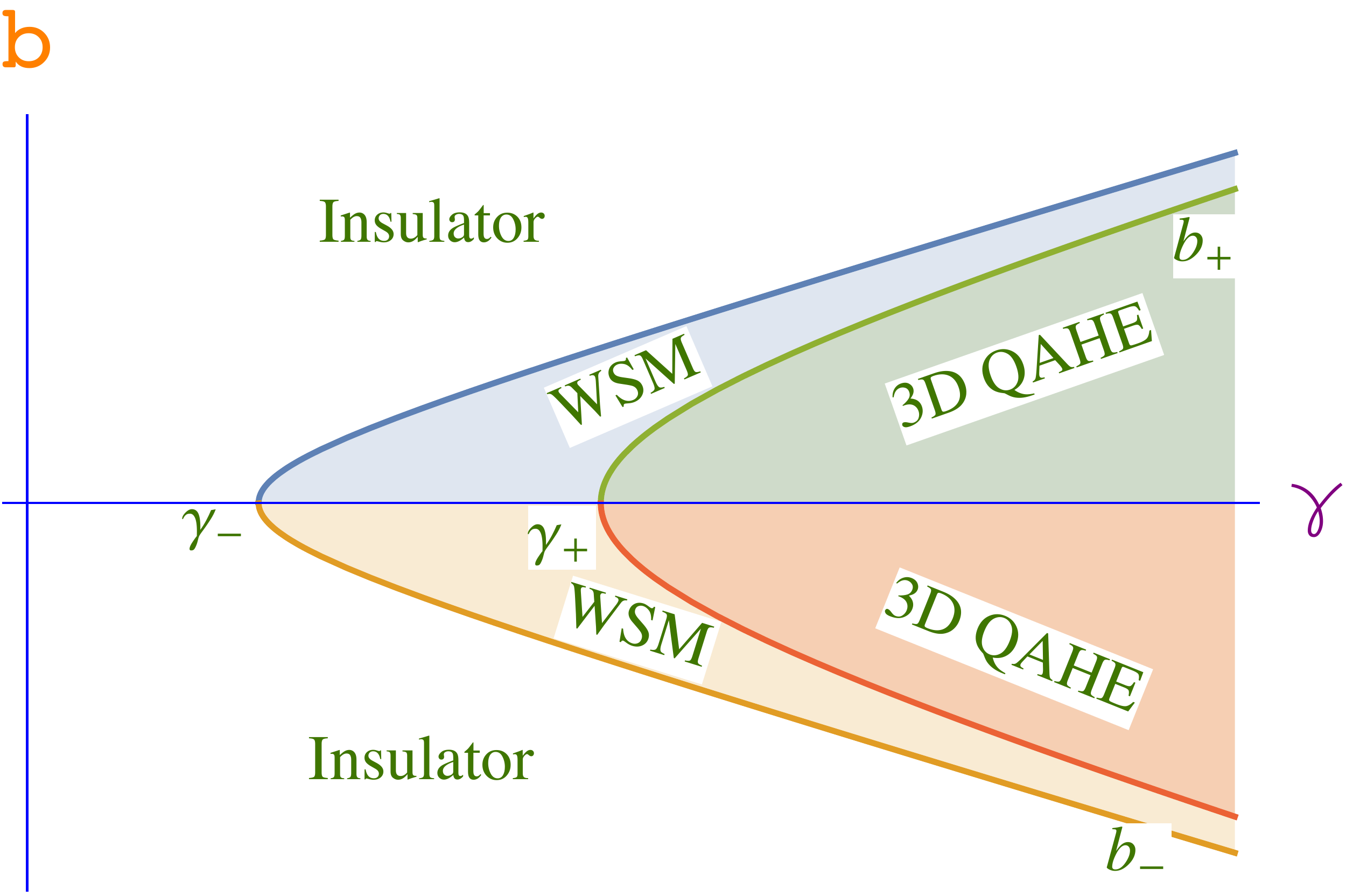}
\caption{Color online. The phase diagram of the thin film heterostructure with the same parameter regimes in Fig.~\eqref{phase}. There are three regimes comprising a Weyl semimetal (WSM), a 3D QAH effect, and an ordinary insulator bounded by the phase boundaries in Eq.~\eqref{bound}.}
\label{phase1}
\end{figure}

 The structure inversion asymmetry breaks  inversion symmetry but preserves time-reversal symmetry (if we take  $\mathcal{T}=\tau_x\otimes \Theta$), whereas the magnetic field breaks time-reversal symmetry and preserves inversion symmetry. The whole system thus breaks both symmetries and  a Weyl semimetal phase is also possible\cite{aab2}.
   The eigenvalues of  Eq.~\eqref{fullti1} are given by
   \begin{align}
&\epsilon_{\eta s}(\bold{k})=\eta\sqrt{[\Delta(k_z, k_\perp)\cos\theta(k_\perp)]^2+m_s^2(k_z,k_\perp)};
\label{bba}
 \end{align}
 where $m_s(k_z,k_\perp)=\xi +s\chi$, 
 \begin{align}
 &\xi=\sqrt{v_F^2k_\perp^2+\gamma^2}; \quad \chi=\sqrt{b^2+\Delta^2\sin^2\theta (k_\perp)};\\&
 \sin\theta(k_\perp)=\frac{\gamma}{\xi};\quad \cos\theta(k_\perp)=\frac{v_Fk_\perp}{\xi}.
 \end{align}
The Weyl nodes are given by the solution of $m_-=0$. We obtain $k_z=\pi/d \pm k_z^0$, where \begin{align}
k_z^0=\frac{1}{d}\arccos\lb 1-\frac{2}{t_D}[\sqrt{\gamma^2-b^2}-\frac{t_S-t_D}{2}]\rb,
\label{nodd1}
\end{align}
and $b<\gamma$. For $b\gg \gamma$, there is neither a circular node nor a Weyl node as can be verified from Eq.~\eqref{bba}.
The phase boundaries are given by
\bea
b_\pm= \eta\sqrt{\gamma^2 -\gamma_\pm^2},
\label{bound}
\eea
where  $b_+<b_-$. The Chern number in Eq.~\eqref{chern} only changes by the phase boundary Eq.~\eqref{bound}.

  Figures~\eqref{phase} and \eqref{phase1} show the energy bands and the phase diagram respectively. Approaching the phase boundaries from the upper bound,  we see that the system is an ordinary insulator  for $b>b_-$. The two Weyl nodes are located at the intermediate regime $b_+ <b<b_-$.  The Weyl nodes, however, occur at the zero energy despite broken time-reversal and inversion symmetries.  At $b=b_+$ they are annihilated at the center of the BZ $k_z^0=\pi/d$ or $k_z=0$.    For  $b<b_+$ the system is  again fully gapped,  which corresponds to  a 3D QAH phase.
\section { Orbital magnetic field effects}
Next, we study the effects of an orbital magnetic field  through the  Peierls substitution, $\bold k_{\perp }\to -i\boldsymbol{\nabla}+\frac{e}{c}\bold{A}$. In the Landau gauge the  vector potential, $\bold{A}=-yB\hat{x}$ and corresponds to a magnetic field along the growth $z$-direction. Introducing the operator $\boldsymbol{\pi}=-i\boldsymbol{\nabla} +\frac{e}{c}\bold{A}$, Eq.~\eqref{fullti} can be written as
\begin{align}
\mathcal H(k_z)= v_F(\pi_y\sigma_x -\pi_x\sigma_y) + [\gamma + \hat{\Delta}(k_z,\pi_\perp)]\sigma_z +b\tau_x,
\end{align}
where \bea
\hat{\Delta}(k_z,\pi_\perp)=[\frac{t_S}{2}+ \frac{t_D}{2}\cos(k_zd)-t_\perp (\pi_x^2+\pi_y^2)]\tau_z.
\eea

The Landau level spectrum  is obtained  by introducing the creation and annihilation operators:
\begin{align}
\pi_x= \frac{a^\dagger +a}{l_B\sqrt{2}}; \quad \pi_y= -i\frac{a^\dagger -a}{l_B\sqrt{2}},
\end{align}
where $l_B^2=c/eB$ is the magnetic length. 
In terms of $a$ and $a^\dagger$ the Hamiltonian becomes
 \begin{align}
\mathcal{H}(k_z)=i\omega_B\sqrt{2}(\sigma^+a-\sigma^-a^{\dagger})+[\gamma + {\Delta}\tau_z]\sigma_z +b\tau_x,
\label{hamm}
\end{align}
 where \begin{align}
{\Delta}(k_z;a^\dagger, a )=\bigg[\frac{t_S}{2}+ \frac{t_D}{2}\cos(k_zd)-\omega_0\lb a^\dagger a+ \frac{1}{2}\rb\bigg].
\end{align}

Here, $\omega_B={v_F}/l_B$ is the magnetic frequency and $\omega_0= 2t_\perp/{l_B^2}$ is the harmonic oscillator frequency. The eigenvalue equation may be written as
\begin{align}
 \begin{pmatrix}
 h_L(k_z) & b \\
b & h_R(k_z)
 \end{pmatrix}
 \begin{pmatrix}\psi_L\\\psi_R\end{pmatrix}=\epsilon(k_z)\begin{pmatrix}\psi_L\\\psi_R\end{pmatrix},
 \label{weylcon}
 \end{align}
 
  where
  \begin{align}
  h_{L/R}(k_z) = i\omega_B\sqrt{2}(\sigma^+a-\sigma^-a^{\dagger})+[\gamma \pm {\Delta}(k_z;a^\dagger, a )]\sigma_z,
  \end{align}
  and $\psi_{L/R}$ are two-component spinors. 
  
  Equation~\eqref{weylcon} shows that the potential $b$ acts as a mass term in $4\times 4$ Dirac equation. When this mass, $b$, vanishes, $\psi_{L}$ and $\psi_{R}$ decouples into two  separate Weyl equations.  For the two coupled Weyl fermions $b\neq 0$, the eigenvectors are spinors given by

  \begin{align}
\psi_{L/R; n\neq 0}=\begin{pmatrix}\alpha_{L/R}u_{n-1}\\\beta_{L/R}u_{n}\end{pmatrix},
\end{align}
where $\alpha_{L/R},\beta_{L/R}$ are constants to be determined. The operators satisfy $au_{n}=\sqrt{n}u_{n-1}$; $a^\dagger u_{n}=\sqrt{n+1}u_{n}$. Hence, Eq.~\eqref{weylcon} yields a secular equation for $\alpha_{L/R}$ and $\beta_{L/R}$

\begin{align}
 \begin{vmatrix}
{\mathcal{R}}_1-\epsilon_n & i\omega_B\sqrt{2n} &b&0 \\
 -i\omega_B\sqrt{2n} & {\mathcal{R}}_2-\epsilon_n &0&b\\
 b &0 & {\mathcal{R}}_3-\epsilon_n& i\omega_B\sqrt{2n}\\
 0 &b & -i\omega_B\sqrt{2n}& {\mathcal{R}}_4-\epsilon_n
 \end{vmatrix}=0,
 \label{dett}
 \end{align}
 where
 \begin{align}
{\mathcal{R}}_{1,2}=\frac{\omega_0}{2}\pm \mathcal{R}_n^+(k_z);~ \mathcal{R}_{3,4}=-\frac{\omega_0}{2}\pm \mathcal{R}_n^-(k_z), 
\label{this}
\end{align}

 \bea
\mathcal{R}_n^s(k_z)=\gamma+s\bigg[\frac{t_S}{2}+ \frac{t_D}{2}\cos(k_zd)-\omega_0 n\bigg].
\eea
The solutions for  $\epsilon_n(k_z)$ correspond to the Landau level spectrum.
%
Next, we drop the zero point energy of the harmonic oscillator in Eq.~\eqref{this}, that is $\omega_0/2$, which can be eliminated by normalizing the oscillator energy. Thus, Eq.~\eqref{dett} is amenable to analytical solution  and the Landau level are given by  

 \begin{align}
&\epsilon_{n,s}^\eta (k_z)=\eta\sqrt{(\xi_n +s\chi_n)^2 +\Delta_n^2\cos^2\theta_n},\quad n\geq 1,\label{eqnn}\\
 &\epsilon^{\eta}_{0}(k_z)=\eta(\gamma -\sqrt{b^2  +\Delta_0^2}), \quad n=0,
 \label{eqnn1}
 \end{align}
 where
 \begin{align}
 &\xi_n=\sqrt{2n\omega_B^2+\gamma^2}; \quad \chi_n=\sqrt{b^2+\Delta_n^2\sin^2\theta_n};\\&
 \sin\theta_n=\frac{\gamma}{\xi_n};\quad \cos\theta_n=\frac{\sqrt{2n}\omega_B}{\xi_n};\\&\Delta_n(k_z)=\frac{t_S}{2}+ \frac{t_D}{2}\cos(k_zd)-\omega_0 n.
 \end{align}
The zero Landau levels in Eq.~\eqref{eqnn1} clearly recover the Weyl nodes derived above.
\section { Continuum limit}
\label{nuum}
Finally, we now consider the continuum limit of the TI thin film multilayer (Eq.~\ref{fullti}). As mentioned above, one of the crucial differences between the present heterostructure and that of Burkov and Balents\cite{aab} is the $\hat{\Delta}(k_z, k_\perp)$ function, which is diagonal in the $\tau$ space.  In the continuum limit we expand the $\hat{\Delta}(k_z, k_\perp)$ function near the Dirac point at $k_x=k_y=0$, $k_z=\pi/d$ and obtain \bea
\hat{\Delta}_c(k_z, k_\perp)=[\frac{t_S-t_D}{2}-t_\perp k_\perp^2+ \frac{\tilde{t}_D}{2}k_z^2]\tau_z,
\label{cont1}
\eea
 with $\tilde{t}_D=d^2t_D/2$ and we have rescaled $k_z\to k_z+\pi/d$.
 The continuum Hamiltonian can then be written as \begin{align}
\mathcal{H}_c(\bold{k})&= v_F(\hat{z}\times \boldsymbol{\sigma})\cdot\bold{k}_\perp +\hat{\Delta}_c(k_z, k_\perp)\sigma_z.
\label{conti}
\end{align}
\begin{figure}[ht]
\centering
\includegraphics[width=2.8in]{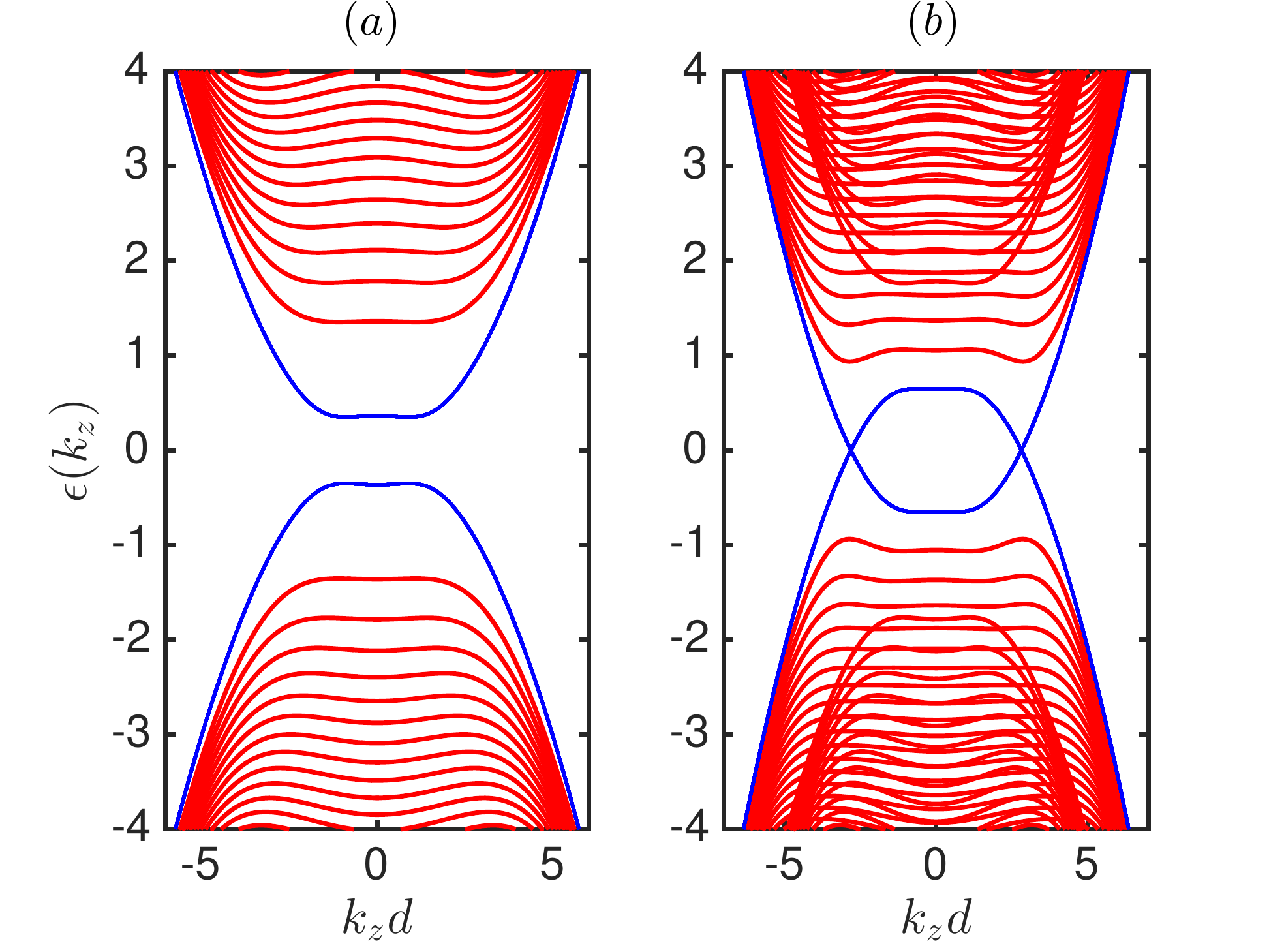}
\caption{Color online. The Landau level energy bands in the continuum limit $(a)~\gamma=0;~b=0.35$; $(b)~\gamma=1,~b=0.35$. All other parameters are $t_D=0.5,~t_S=0.8t_D,~t_\perp=0.1, v_F=g=B=1$. The universal constants $e,~c,~\mu_B$ are set to unity. The blue curves denote the zero Landau levels.}
\label{LLs}
\end{figure}

  Using Eq.~\eqref{conti}, it is readily seen that the Dirac nodes in the continuum limit  are located at $\bold{k}^c=(0,0,\pm k_z^c)$, where
 \begin{align}
 k_z^c= \frac{1}{d}\sqrt{2(1-t_S/t_D)}; 
 \end{align}
provided $t_S/t_D<1$. Similar expression can be found for the general case $b,\gamma\neq 0$. 
 Near the Dirac points, the dispersion is that of a massless 3D Dirac equation, given by
\begin{align}
\epsilon_\pm(\bold q)\approx \pm\sqrt{v_F^2q_x^2 +v_F^2 q_y^2+ \tilde{v}_F^2 q_z^2},
\end{align}
where $\tilde{v}_F=\tilde{t}_D k_z^c$ and $\bold{q}=\bold{k}-\bold{k}^c$ is the momentum deviation from the Dirac nodes. Figure~\eqref{LLs} shows the Landau levels in the continuum limit for $\gamma=0,~b\neq 0$ (a) and $\gamma\neq 0,~b\neq 0$ (b). The Landau levels capture the appearance of Weyl (nodal) semimetal.  We find no evidence of Weyl nodes  for $\gamma=0,~b\neq 0$  as can be seen  from the phase boundary in Eq.~\eqref{bound}.
Each pseudo spin sector $(\tau_z=\pm 1)$  in Eq.~\eqref{cont1} is the same as the continuum limit of the toy tight binding Hamiltonian of Weyl semimetal \cite{hui5, hui6,hui7,hui8}. 

\section { Conclusion}
In this paper, we have studied a Weyl semimetal model first introduced by Burkov and Balents\cite{aab}, but with an ultrathin film of topological insulator layers, whose Hamiltonian is different from that of the surface state of a strong 3D  topological insulator. In this new model, a 3D quantum spin Hall state emerged and the nodes of the Weyl semimetal occur at zero energy when time-reversal and inversion symmetries are broken. This new heterostructure  can also be grown in the laboratory using ultrathin Bi$_2$Se$_3$ and Bi$_2$Te$_3$ films. Hence, it is experimentally accessible. Another important feature of the present model is that it reproduces previously studied toy tight binding models\cite{hui5, hui6,hui7, hui8} of Weyl semimetal in the continuum limit. Thus, it offers a physical realization  of a Weyl semimetal phase in those systems.   Thus, the ultra-thin film of topological insulator   multilayer is another candidate for realizing Weyl semimetal phases (nodal semimetal) and quantum spin Hall state in three-dimensional electron systems.

\section*{ Acknowledgement}
 We thank African Institute for Mathematical Sciences for hospitality. Research at Perimeter Institute is supported by the Government of Canada through Industry Canada and by the Province of Ontario through the Ministry of Research
and Innovation.


\begin{thebibliography}{99}
\bibitem{aab}
A. A. Burkov and L. Balents, \prl {\bf 107}, 127205 (2011).
\bibitem{aab0}
 A. A. Burkov, M. D. Hook, and L. Balents, \prb {\bf 84}, 235126 (2011).
 \bibitem{gab}
 G. B. Halász and L. Balents,  \prb {\bf 85}, 035103 (2012).
  \bibitem{mur}
S. Murakami, New J. Phys. {\bf 9}, 356 (2007).
\bibitem{wan}
X. Wan {\it et al}., \prb {\bf 83}, 205101 (2011).
\bibitem{zhang1}
X. -L.  Qi, T.  L. Hughes, and S. -C.  Zhang, 
\prb {\bf 78}, 195424 (2008); \prb {\bf 81}, 159901  (2010).
\bibitem{zhang}
X. -L.  Qi and S. -C.  Zhang, \rmp{\bf 83}, 1057 (2011).
\bibitem{zhang0}
 H. Zhang, {\it et al.}, Nature Phys. {\bf 5}, 438 (2009). L.  Wu, {\it et al.} Nature Physics {\bf 9}, 410, 2013.
\bibitem{hk}
M. Z. Hasan and C. L. Kane, \rmp{\bf 82}, 3045 (2010); 
 C. L. Kane and E. J. Mele, \prl {\bf 95}, 226801 (2005).
 \bibitem{batt}
B. A. Bernevig, T. L. Hughes, and S. -C. Zhang, Science {\bf 314}, 1757 (2006). M.  Koenig, {\it et al.}, J. Phys. Soc. Jpn. {\bf 77}, 031007 (2008).
\bibitem{vol}
F. R. Klinkhamer, G. E. Volovik, Int. J. Mod. Phys. {\bf A20}, 2795  (2005) ; G. E. Volovik, The Universe in a Helium Droplet, Oxford University Press, (2003).
\bibitem{solu}
A. A. Soluyanov, {\it et al}., Nature {\bf 527}, 495 (2015).

\bibitem{kre}
W. Witczak-Krempa and Y. B. Kim, \prb
{\bf 85}, 045124 (2012).
\bibitem{aab1}
A. A. Zyuzin, M. D. Hook, A. A. Burkov, \prb {\bf 83}, 245428 (2011).
\bibitem{aab2}
A. A. Zyuzin, S. Wu, and A. A. Burkov, \prb {\bf 85}, 165110 (2012).
\bibitem{liu}
C. -X. Liu, P. Ye, X. -L. Qi,  \prb {\bf 87}, 235306 (2013); G. Y. Cho, arXiv:1110.1939 [cond-mat.str-el].
\bibitem{hui5}
K.-Y. Yang, {\it et al.}, \prb {\bf 84}, 075129 (2011).
\bibitem{hui6}
C. -Z. Chen, {\it et al.}, \prl {\bf 115}, 246603 (2015).
\bibitem{hui7}
H. -Z. Lu, S. -B. Zhang, and S. -Q. Shen, \prb {\bf 92}, 045203 (2015).
\bibitem{hui8}
 P.  Delplace, J.  Li, D.  Carpentier,  EPL {\bf 97}, 67004 (2012).



\bibitem{llu}
L. Lu, {\it et al}., Science, {\bf 349}, 622 (2015).
\bibitem{Xu}
S. -Y. Xu, {\it et al}., Science, {\bf 349}, 613 (2015).
\bibitem{lv}
B. Q. Lv, {\it et al}., Phys. Rev. X {\bf 5}, 031013 (2015).
\bibitem{lv1}
B. Q. Lv, {\it et al}., Nature Physics {\bf 11}, 724 (2015).


\bibitem{hai1}
H. -Z Lu, {\it et al.},  \prb {\bf 81}, 115407 (2010).
\bibitem{hui}
H.  Li, {\it et al.}, \prb {\bf 82}, 165104,  (2010).
\bibitem{hui1}
H.  Li, L. Sheng, and D. Y. Xing, \prb {\bf 85}, 045118 2012;  \prb {\bf 84}, 035310 (2012).
\bibitem{hui2}
 W. -Y. Shan, H. -Z. Lu, and S. -Q. Shen, New J. Phys. {\bf 12}, 043048
(2010).
 \bibitem{zza1}
Y.  Sakamoto {\it et al}., \prb {\bf 81}, 165432 (2010).

\bibitem{hui4}
Y.  Zhang, {\it et al.}, Nature, {\bf 6}, 584 (2010).
\bibitem{hui3}
J. Wang, {\it et al.}, \prb {\bf 83}, 245438 (2011).
\bibitem{luz}
S. -B. Zhang, H. -Z. Lu, S. -Q. Shen, Scientific Reports {\bf 5}, 13277 (2015).
\bibitem{yang}
B. -J. Yang, T.  Morimoto, and A.  Furusaki, \prb {\bf 92}, 165120 (2015). 
 \bibitem{wang1}
Z.  Wang, {\it et al}., \prb{\bf 85}, 195320 (2012).
\bibitem{wang2}
Z.  Wang, {\it et al}., \prb{\bf 88}, 125427 (2013).
\bibitem{wang}
S. M. Young, {\it et al}., \prb{\bf 108}, 140405 (2012).


\bibitem{fuk1}
L. Fu and C. L. Kane, \prb {\bf 76}, 045302 (2007).
\bibitem{fuk}
 S. A. Owerre, and J. Nsofini, Solid State Commun.  {\bf 218}, 35  (2015).
\end{thebibliography}
\end{document}